\documentclass[fleqn,twoside]{article}
\usepackage{espcrc2}

\usepackage{graphicx}

\newcommand{\AmS}{{\protect\the\textfont2
  A\kern-.1667em\lower.5ex\hbox{M}\kern-.125emS}}

\newcommand{\be}{\begin{equation}}
\newcommand{\ee}{\end{equation}}
\newcommand{\beq}{\begin{eqnarray}}
\newcommand{\eeq}{\end{eqnarray}}

\title{ Gauge-invariant two- and three- density
correlators\thanks{Talk presented by C.~Alexandrou.}}

\author{C. Alexandrou\address{Department of Physics, University of Cyprus,
CY-1678 Nicosia, Cyprus},
Ph.\ de Forcrand\address{ETH-Z\"urich, CH-8093 Z\"urich and CERN Theory Division, CH-1211 Geneva 23, Switzerland} 
and A. Tsapalis\address{Department of Physics, University of Wuppertal, Wuppertal, Germany} }
       
\begin{document}

\begin{abstract}
Gauge-invariant spatial correlations between two and three quarks 
inside a hadron are measured
within quenched and unquenched QCD. These correlators
provide information on the shape and multipole moments of the pion, the rho,
the nucleon and the $\Delta$.
\vspace{1pc}
\end{abstract}

\maketitle

\vspace*{-1.8cm}

\section{Introduction}

 Gauge-invariant two- and three-density correlators inside a hadron reduce to the square of the wave function in the non-relativistic limit, yielding detailed information
on hadron structure.
Quark spatial distributions, hadronic shapes, charge radii,
etc., can be extracted. 
An interesting question is whether the nucleon is deformed.
 Strong evidence for deformation in the nucleon and/or $\Delta$ is provided by recent accurate measurements~\cite{Athens} 
 in  photoprodution  experiments on the nucleon.
One can compare with the experimental results by 
calculating the $N$ to $\Delta^+$ transition form factors~\cite{antonis}.
Moreover,
the $\Delta^+$ deformation can be directly obtained from the wave function.
The nucleon, however, has zero spectroscopic quadrupole moment,
since it is a spin $1/2$ particle.
This implies that any deformation that intrinsically may be present for a
given background gauge field will average out to zero
in the gauge configuration ensemble.

\vspace{-0.3cm}

\section{Observables}
In this work we consider
equal-time correlators. For a meson we calculate the matrix element
\be
 C_\Gamma({\bf r},t) = \int\> d^3r'\>
\langle M|\rho^u({\bf r'},t)\rho^{d}({\bf r}'+{\bf r},t)|M\rangle 
\ee
with
$\rho^u_\Gamma({\bf r},t)=:\bar{u}({\bf r},t)\Gamma u({\bf r},t):$.
We take $\Gamma=\gamma_0$ for 
 the normal ordered charge density operator, and ${\bf 1}$ for 
the matter density. 
In the non-relativistic limit this allows to extract respectively
the charge and matter density distributions.
Unlike Bethe-Salpeter amplitudes, these correlators are gauge-invariant.
Here we take the $u- $and $d-$ quarks to be degenerate in mass.

The form factors $F({\bf q^2})$ at low momenta can  be extracted from the Fourier transform of
the charge density-density correlator~\cite{Wilcox}

\vspace{-0.5cm}

\beq
Q({\bf q}^2) &= &\int_0^\infty d^3r\>\exp{(i{\bf q.r})}\> C_{\gamma_0}({\bf r},t)/2m_\pi \nonumber \\ 
&\rightarrow&
\frac{(E+m_\pi)^2}{4Em_\pi} F({\bf q^2})^2 \quad.
\eeq

\vspace{-0.2cm}

\begin{figure}[t]
\begin{center}
\mbox{\includegraphics[height=4cm]{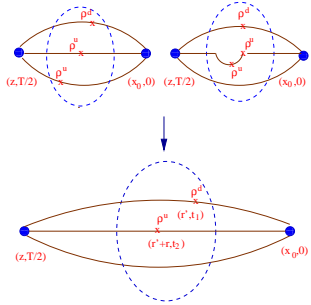}}
\vspace*{-0.8cm}
\caption{Three-density correlator for a baryon where all the insertions
are taken at equal times $t$.
$t$ and $T/2-t$ are taken large enough to filter out the excited baryonic states.}
\end{center}
\label{fig:baryons}
\vspace*{-0.4cm}
\end{figure}

For baryons the charge distribution is obtained by using 
three density insertions. This  involves two relative distances and it is
 computed efficiently by using  FFT.
The diagrams involved are shown in Fig.~\ref{fig:baryons}.
The top right diagram requires the all-to-all propagator~\cite{AFT}
and will not be considered here.
We
check that the contribution of this diagram is small
by comparing  the integrated one-particle density 
$\int d^3 r_2 C({\bf r}_1,{\bf r}_2,t)$
 with the two-density correlator, $\int\> d^3r'\>\langle h|\rho^{d}({\bf r}',t)\rho^{u}({\bf r}'+{\bf r}_1,t)|h\rangle$, shown by  
the lower diagram of Fig.~\ref{fig:baryons}.

\vspace*{-0.3cm}

\section{Results}
We first discuss the results obtained in the quenched approximation 
at  $\beta=6.0$ for a lattice of size $16^3\times32$. The parameters
of the simulation for both quenched and unquenched calculations
are summarized in the following Table.

\begin{table}
\small
\vspace*{-0.8cm}
\begin{tabular}{|ccc|ccc|}\hline 
\multicolumn{3}{|c|}{Quenched}& \multicolumn{3}{|c|}{Unquenched(from~\cite{SESAM})}\\ \hline
  $\kappa$ & $m_\pi/m_\rho$& \#confs.& $\kappa_{sea}$ &$m_\pi/m_\rho$ & \#confs. \\ \hline
 0.153  &0.81 & 220& 0.156  & 0.83 &150\\
 0.154  & 0.78 & 220& 0.157  & 0.76 &200\\ \hline 
\end{tabular}
\vspace*{-0.8cm}
\end{table}
\normalsize

\begin{figure}[t]
\mbox{\includegraphics[height=4cm,width=6cm]{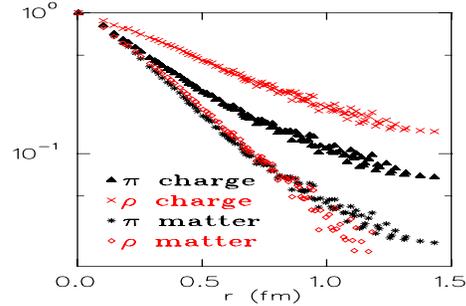}}
\vspace*{-1cm}
\caption{Charge and matter density distributions at $\kappa=0.154$
for the pion and the rho.}
\vspace*{-0.75cm}
\label{fig:ch_matter}
\end{figure}

\begin{figure}[tb]
\mbox{\includegraphics[height=4.cm,width=6cm]{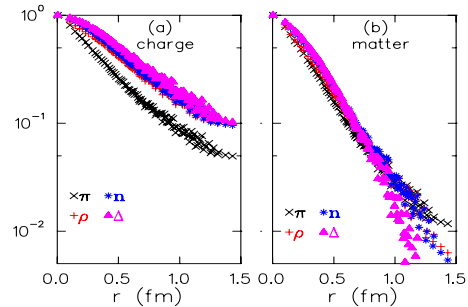}}
\vspace*{-1cm}
\caption{Charge and matter density distributions at $\kappa=0.153$.}
\label{fig:hadrons_matter}
\vspace*{-0.8cm}
\end{figure}

The charge and matter density distributions for the pion 
and the rho mesons are
shown in Fig.~\ref{fig:ch_matter} at $\kappa=0.154$.
We find that the charge distribution is broader than the matter distribution,
in agreement with  Ref.~\cite{Green}. The 
matter distributions for the pion and the rho  are almost identical,
but the density distribution is considerably
broader for the rho.
As seen in  Fig.~\ref{fig:hadrons_matter},
the nucleon and the $\Delta^+$ 
 have very similar matter distributions to the rho
meson, whereas larger variations are visible in the charge density distributions.

\begin{figure}[t]
\begin{center}
\mbox{\includegraphics[height=4cm,width=6cm]{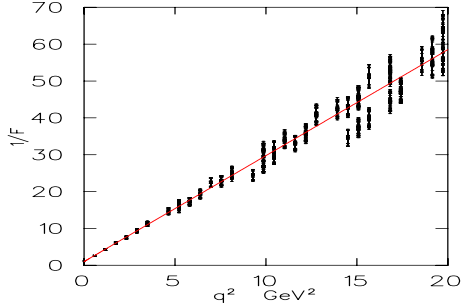}}
\end{center}
\vspace*{-1cm}
\caption{$1/F({\bf q}^2)$ versus ${\bf q}^2$
 at $\kappa=0.154$. The
solid line is a fit to the form $(1+{\bf q}^2/m^2)$.}
\vspace*{-0.5cm}
\label{fig:pion form factor}
\end{figure}

 In Fig.~\ref{fig:pion form factor} we show the pion form factor
which, being the simplest, can be reliably extracted  from
our set of data.
Its low momentum behaviour is well described by the vector dominance 
 result $F({\bf q}^2) \sim 1/(1+{\bf q}^2/m^2)$.
We observe a weak dependence on $\kappa$ obtaining 0.57, 0.59 and 0.61 GeV
at $\kappa=0.153,0.1554$ and 0.155 respectively. Extrapolating linearly in $m_\pi^2$
we obtain $m \sim 0.65$~GeV within 15\% of the rho meson mass.

\begin{figure}[tb]
\mbox{\includegraphics[height=5cm,width=6cm]{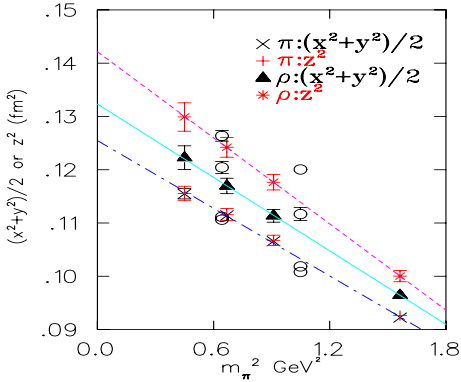}}
\vspace*{-1cm}
\caption{$\langle z^2 \rangle $ and $\langle (x^2+y^2)/2 \rangle $ vs
the pion mass squared. The lines are linear fits to the
quenched data. The empty circles show full QCD results.} 
\label{fig:radii}
\vspace*{-0.7cm}
\end{figure}

Hadron deformation can be investigated by 
evaluating  the root mean squared (rms) 
radius along the spin axis and perpendicular to it.
We show the charge rms radii in Fig.~\ref{fig:radii} as a  function of the
quark mass. The pion, having equal rms radii, is spherical whereas for the rho 
the asymmetry between the longitudinal and transverse 
radii  increases as we approach the 
chiral limit. A reliable extrapolation to the chiral limit
 needs lighter quarks and a larger lattice to contain the rho.
  A similar analysis for the nucleon shows that it is spherical
as expected, whereas the $\Delta^+$ shows no statistically significant
deformation~\cite{AFT}.
The rms radii computed from the matter distribution show no such deformation
also in the case of the rho. 

\normalsize
\begin{figure}[t]
\vspace*{-0.5cm}
\centerline{\mbox{\includegraphics[height=4cm,width=6cm]{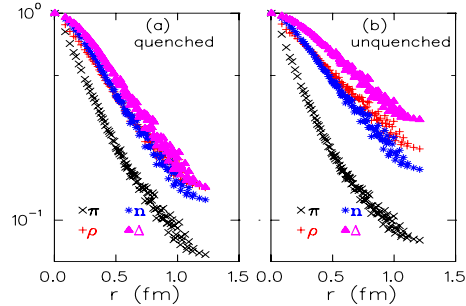}}}
\vspace*{-1cm}
\caption{(a) Density-density correlators, $ C_{\gamma_0}({\bf r})$,
for the pion, the rho, the nucleon and
the $\Delta^+$ at $\kappa=0.154$ vs $|{\bf r}|$.
(b) Same as (a) but with two dynamical quark flavors at $\kappa=0.157$.
Errors bars are omitted for clarity.}
\vspace*{-0.5cm}
\label{fig:hadronwfs}
\end{figure}

\begin{figure}[tb]
\mbox{\includegraphics[height=4cm,width=6cm]{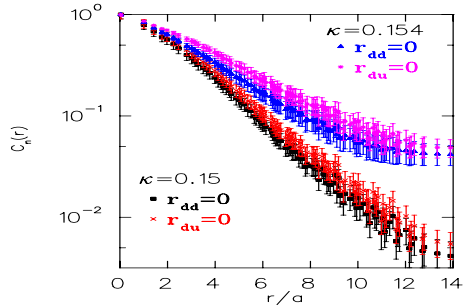}}
\vspace*{-1cm}
\caption{The $u-$ and $d-$ distributions inside the neutron for $\kappa=0.15$ and 0.154.}
\vspace*{-0.7cm}
\label{fig:neutron}
\end{figure}

We now compare our quenched results for the charge
correlator with those obtained using
two degenerate flavors of Wilson fermions, with similar lattice spacing and
$m_\pi/m_\rho$ values. 
Fig.~\ref{fig:hadronwfs} shows that unquenching leaves the pion unchanged,
whereas the rho, the nucleon and the $\Delta^+$ become broader.
The asymmetry in the rho increases as seen by comparing the
 longitudinal and
transverse radii in Fig.~\ref{fig:radii}.
The $\Delta^+$ shows a  small asymmetry which points to pion cloud 
contributions to the deformation.

The three-density charge correlator was computed in the quenched approximation
using 30 configurations at $\kappa=0.15$ and 0.154. More detailed information
on hadron structure can be extracted from this correlator. As
an example we show in Fig.~\ref{fig:neutron} the $u$- and  $d$- quark 
spatial distributions. The systematically broader
$d$-quark distribution  gives rise to a  negative neutron rms charge radius.

\vspace*{-0.3cm}

\section{Conclusions}
We have presented a  gauge-invariant determination of hadron charge and matter
density distributions. The main conclusions from this study are the
following: 
1) The matter density distribution  is very similar for the $\pi, \rho$, nucleon and
$\Delta^+$,
 unlike the charge density which is narrower for the pion
than for the rest. In all cases the  charge density distribution 
is  broader
than the matter density distribution.
 2) Vector dominance provides
a good description of the pion form factor at low momenta. 
3) Unquenching leaves  the pion size  unchanged but the 
rho, the nucleon and $\Delta^+$ become broader.
4) In the quenched approximation the rho in the  0-spin projection
is prolate 
whereas the $\Delta^+$ has no statistically significant 
deformation.
5) Unquenching 
 increases the rho deformation
and produces a small deformation  for the  $\Delta^+$ with the 
$+3/2$ spin state 
being slightly oblate.
6) The $d$-quark spatial distribution in the neutron 
extracted from the three density correlator is broader
than that of the $u$-quark, thus accounting for the negative rms charge 
radius of the neutron.


\vspace*{-0.3cm}

\begin{thebibliography}{99}
\bibitem{Athens}C.~Mertz {\it et al.},
Phys.\ Rev.\ Lett.\  {\bf 86} (2001) 2963; K. Joo {\it et al.}, 
Phys.\ Rev.\ Lett.\ {\bf 88} (2002) 122001.
\bibitem{antonis} See talk by A.~Tsapalis, C.~Alexandrou {\it et al.}, these proceedings.
\bibitem{Wilcox}
W.~Wilcox,
Phys.\ Rev.\ D {\bf 43}, 2443 (1991).
\bibitem{AFT} C.~Alexandrou, Ph.~de Forcrand and A.~Tsapalis, hep-lat/0206026.
\bibitem{SESAM} 
N.~Eicker {\it et al.}, 
Phys.\ Rev.\ D {\bf 59} (1999) 014509.
\bibitem{Green} A. Green, J. Koponen, P. Pennanen, C. Michael, hep-lat/0206015.
\end{thebibliography}
\end{document}